\documentclass[aip, rsi, reprint, twocloumn]{revtex4-1}
\usepackage[pdftex]{graphicx}
\usepackage{color,amsmath,amssymb}

\begin{document}
  \title{Interaction between Faraday rotation and Cotton-Mouton effects in polarimetry modeling for NSTX}

  \author{J. Zhang}
  \email{xyzhangj@physics.ucla.edu}
  \author{N. A. Crocker}
  \author{T. A. Carter}
  \author{S. Kubota}
  \author{W. A. Peebles}
  \affiliation{UCLA Physics and Astronomy Department, Los Angeles, California 90095-1547}
  \author{and the NSTX research Team}
  \affiliation{Princeton Plasma Physics Laboratory, Princeton, New Jersey 08543-0451}

  \begin{abstract}
The evolution of electromagnetic wave polarization is modeled for propagation in the major radial direction in the National Spherical Torus Experiment (NSTX) with retroreflection from the center stack of the vacuum vessel. This modeling illustrates that the Cotton-Mouton effect--elliptization due to the magnetic field perpendicular to the propagation direction--is shown to be strongly weighted to the high-field region of the plasma. An interaction between the Faraday rotation and Cotton-Mouton effects is also clearly identified. Elliptization occurs when the wave polarization direction is neither parallel nor perpendicular to the local transverse magnetic field. Since Faraday rotation modifies the polarization direction during propagation, it must also affect the resultant elliptization. The Cotton-Mouton effect also intrinsically results in rotation of the polarization direction, but this effect is less significant in the plasma conditions modeled. The interaction increases at longer wavelength, and complicates interpretation of polarimetry measurements.
  \end{abstract}

\maketitle

\graphicspath{{Figures/}}

\section{Introduction}
Polarimetry is a powerful technique for probing magnetic field equilibria and fluctuations, plasma density and current density profiles in magnetically confined plasmas.\cite{soltwisch_current_1992, brower_multichannel_2001} It measures changes in the electromagnetic (EM) wave polarization caused by propagation through a magnetized plasma. It is routinely used on conventional high aspect ratio tokamaks (e.g.\ JET\cite{boboc_simultaneous_2006}) and reversed field pinches (RFP, e.g.\ MST\cite{ding_measurement_2003}). However, no detailed study of polarimetry has been performed for propagation in the major radial direction in spherical tori. In contrast with conventional tokamaks and RFPs, in spherical tori both magnetic field strength and direction vary strongly in the major radial direction. The modeling is motivated to guide the design of a polarimeter system planned for the National Spherical Torus Experiment (NSTX)\cite{ono_exploration_2000}. It calculates the evolution of EM wave polarization in the major radial direction with retroreflection from the center stack of the vacuum vessel. The Cotton-Mouton effect--elliptization due to the magnetic field perpendicular to the propagation direction--is shown to be strongly weighted to the high-field region in NSTX. An interaction between the Faraday rotation and Cotton-Mouton effects is also clearly identified. Elliptization occurs when the wave polarization direction is neither parallel nor perpendicular to the local transverse magnetic field. Since Faraday rotation modifies the polarization direction during propagation, it must also affect the resultant elliptization. The Cotton-Mouton effect also intrinsically results in rotation of the polarization direction, but this effect is less significant in the plasma conditions modeled. The interaction is shown to increase with wavelength.

The interaction is present when the magnetic field has both parallel and perpendicular components with respect to the wave propagation direction. It complicates the interpretation of polarimetry measurements, especially at longer wavelength. Previous polarimetry studies focused separately on Faraday rotation\cite{soltwisch_current_1986,zilli_realization_2000,van_zeeland_2008} or the Cotton-Mouton effect\cite{fuchs_cotton-mouton_1998}. Recent results combining measurement and modeling including both effects on JET also assume one effect or the other is small.\cite{boboc_simultaneous_2006,orsitto_modelling_2008} However, this will not be the case in ITER, where both effects are large.\cite{donne_poloidal_1999}

In the following sections, the polarimetry model is described and results are shown for modeling using a plasma density profile from Thomson scattering measurement and a magnetic field profile from EFIT in NSTX.

\section{Polarimetry modeling description}
Several assumptions are made to simplify the calculation of polarization evolution. A cold plasma model is adopted, which excludes corrections from finite temperature effects\cite{mirnov_finite_2007}. The plasma is assumed to be collisionless, so the beam experiences no dissipation. The WKB approximation\cite{swanson_plasma_2003} is used, i.e.\ plasma parameters are assumed to be slowly varying ($|\vec{B}|\gg|(1/k)(\partial{\vec{B}}/\partial{z})|$,$n\gg|(1/k)(\partial{n}/\partial{z})|$), so cutoffs and resonances are not considered. Only the electron response is included; the contribution from ion motion is ignored ($\omega_{pi},\omega_{ci}\ll\omega_{pe},\omega_{ce}\ll\omega$, where $\omega$ is the circular frequency of the probing beam, $\omega_p = \sqrt{n q^2 / m \varepsilon_0}$, $\omega_c = |q|B/m$, subscripts `$i$' and `$e$' stand for ions and electrons respectively). Refraction is also neglected, so it is assumed that the beam path through the plasma is straight and that the beam neither diverges nor converges.

\begin{figure}
\includegraphics[width = 8.5 cm]{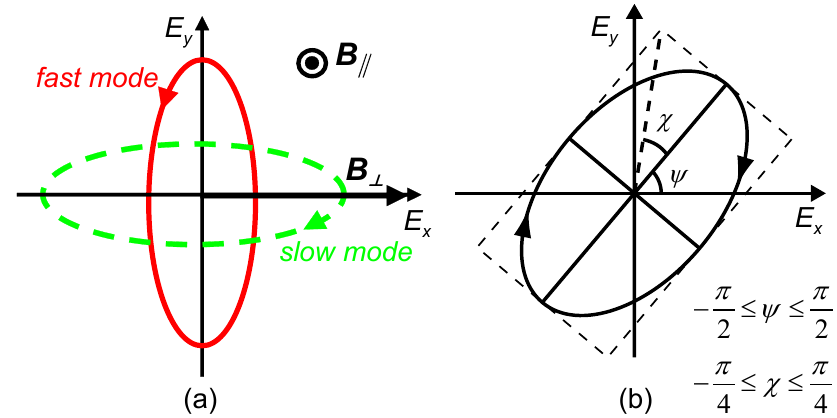}
 \caption{\label{fig:ellipse}(a) {\color{red}Fast (solid line)} and {\color{green}slow (dashed line)} characteristic modes. The fast mode has a major axis perpendicular to $\vec{B_{\perp}}$ and right-handedness with respect to $\vec{B_{\parallel}}$. `$\perp$' and `$\parallel$' are defined with respect to the propagation direction. (b) Polarization properties are characterized by $\chi$ (\textit{elliptization angle}, where right/left handedness are represented by $+/-$ sign) and $\psi$ (\textit{polarization direction angle}); their ranges are also shown.}
\end{figure}

The polarization of an EM wave evolves as it propagates through a magnetized plasma due to plasma birefringence. The plasma features a pair of fast and slow characteristic modes (i.e.\ EM waves that propagate with their polarizations unchanged) whose phase velocities and polarizations are determined by local plasma parameters at any position along the wave path [Fig.~\ref{fig:ellipse}(a)]. The two modes are generally elliptically polarized with orthogonal polarization directions and opposite handedness. The fast mode has a major axis perpendicular to $\vec{B_{\perp}}$ and right-handedness with respect to $\vec{B_{\parallel}}$. (`$\perp$' and `$\parallel$' are defined with respect to the propagation direction) An EM wave of any polarization may be represented as some combination of this pair of characteristic modes. The combined polarization is sensitive to the relative phase of its two components, so a difference in their phase velocities causes the polarization to evolve. The Faraday rotation and Cotton-Mouton effects are two well-known special cases. In Faraday rotation, where the wave propagates parallel to a magnetic field, the characteristic modes are circularly polarized. For the Cotton-Mouton effect, where the wave propagates perpendicular to a magnetic field, the fast and slow modes, which are linearly polarized, are extraordinary and ordinary modes respectively.

The modeling uses the Mueller-Stokes calculus\cite{segre_review_1999}. The ellipse representing the polarization of a single-frequency EM wave is characterized by two parameters, elliptization angle, $\chi$, and polarization direction angle, $\psi$. [Fig.~\ref{fig:ellipse}(b)] The polarization state of the wave can be mapped by the \textit{Stokes vector} $\vec{s}$ (Eq.~\ref{eq:s}) to a point on a unit sphere in an abstract space referred to as the \textit{Poincar$\acute{\textrm{e}}$ sphere}. For instance, alignment of $\vec{s}$ with the $s_3$ axis corresponds to circular polarization, while a vanishing $s_3$ component corresponds to linear polarization in the laboratory frame. As a wave propagates through a magnetized plasma, its polarization evolves and the corresponding Stokes vector traces out a trajectory on the Poincar$\acute{\textrm{e}}$ sphere. Each small step of the trajectory results from a small rotation of the Stokes vector around an axis given by the vector $\vec{\Omega}$ which is determined everywhere along the wave path by local plasma parameters. (Eq.~\ref{eq:rotation}) The $z$ coordinate indicates position along the wave path and $c$ is speed of light in vacuum.
\begin{equation}
\label{eq:s} \vec{s}=  \left(
  \begin{array}{c}
  s_1\\
  s_2\\
  s_3
  \end{array}\right)
  =  \left(
  \begin{array}{c}
  \cos 2\chi\cos 2\psi\\
  \cos 2\chi\sin 2\psi\\
  \sin 2\chi
  \end{array}\right)
\end{equation}
\begin{equation}
  \label{eq:rotation}
  \frac{\mathrm{d}{}\vec{s} (z)}{\mathrm{d}{} z}=\vec{\Omega}(z)\times\vec{s}(z),
  \vec{\Omega}=\frac{\omega_{pe}^2 \omega_{ce}^2}{2c \omega (\omega^2-\omega_{ce}^2)}
  \left (\begin{array}{c}
  (B_{x}^{2}-B_{y}^{2})/B^2\\
  2B_{x}B_{y}/B^2\\
  2(\frac{\omega}{\omega_{ce}})B_{z}/B
  \end{array} \right)
 \end{equation}
The origin of the interaction between the Faraday rotation and Cotton-Mouton effects can be seen clearly from the preceding geometrical description of the polarization evolution. Both the Faraday rotation and Cotton-Mouton effects are directly related to the components of $\vec{\Omega}$. A non-vanishing $\Omega_{3}$ gives rise to Faraday rotation by causing a rotation of $\vec{s}$ about the $s_3$ axis. This corresponds to a rotation of the wave polarization ellipse in the laboratory frame. A non-vanishing $\Omega_{1}$ or $\Omega_{2}$ gives rise to the Cotton-Mouton effect by causing a change in $s_3$ and therefore in the ellipticity of the wave polarization ellipse. However, the way in which $s_3$ changes clearly depends on the direction of $\vec{s}$ relative to $\vec{\Omega}$, which can be influenced by Faraday rotation.

The interaction may also be seen from the differential expressions relating changes in $\chi$ and $\psi$ to the Faraday rotation (FR) and Cotton-Mouton (CM) effects\cite{guenther_approximate_2004}:
\begin{subequations}
  \label{eq:chipsi}
  \begin{equation}
    \label{eq:chi}
    \mathrm{d} \chi = \frac{1}{2}\sin 2\psi \mathrm{d} \delta |_{\mathrm{CM}}
    \end{equation}
    \begin{equation}
    \label{eq:psi}
    \mathrm{d} \psi = \mathrm{d}\psi |_{\mathrm{FR}}-\frac{1}{2}\tan 2\chi \cos 2\psi \mathrm{d} \delta |_{\mathrm{CM}}
    \end{equation}
\end{subequations}
\begin{subequations}
  \label{eq:chipsi_app}
    \begin{equation}
    \label{eq:psi_app}
    \mathrm{d}\psi |_{\mathrm{FR}} = -\frac{\omega_{pe}^2 \omega_{ce}}{2c\omega^2} \frac{B_{\parallel}}{B} \mathrm{d}z
    \end{equation}
    \begin{equation}
    \label{eq:chi_app}
    \mathrm{d} \delta |_{\mathrm{CM}} = \frac{\omega_{pe}^2 \omega_{ce}^2}{2c\omega^3} (\frac{B_{\perp}}{B})^2 \mathrm{d}z
    \end{equation}
\end{subequations}
 $\delta$ is the relative phase between the $x$ and $y$ components of the wave electric field. Eq.~\ref{eq:chipsi} assumes the coordinate system illustrated in Fig.~\ref{fig:ellipse}(b) where the $x$ axis is aligned with $\vec{B_{\perp}}$. Eq.~\ref{eq:chi} shows the sensitivity of elliptization to $\psi$. Faraday rotation modifies $\psi$, so it affects elliptization. Eq.~\ref{eq:psi} shows that the Cotton-Mouton effect also intrinsically causes polarization rotation, but for the cases modeled here, this contribution to the total polarization rotation proves to be relatively small. This is because through out the majority of the wave path, either the absolute elliptization angle is small (i.e.\ $|\chi| \ll 45^{\circ}$) or the Cotton-Mouton effect is overwhelmed by Faraday rotation (i.e.\ $|\mathrm{d}\delta |_{\mathrm{CM}}|\ll |\mathrm{d}\psi |_{\mathrm{FR}}|$). The following discussion will focus on the impact of Faraday rotation on elliptization.

\section{Modeling Results and Discussion}
\begin{figure}
\includegraphics[width = 8.5 cm]{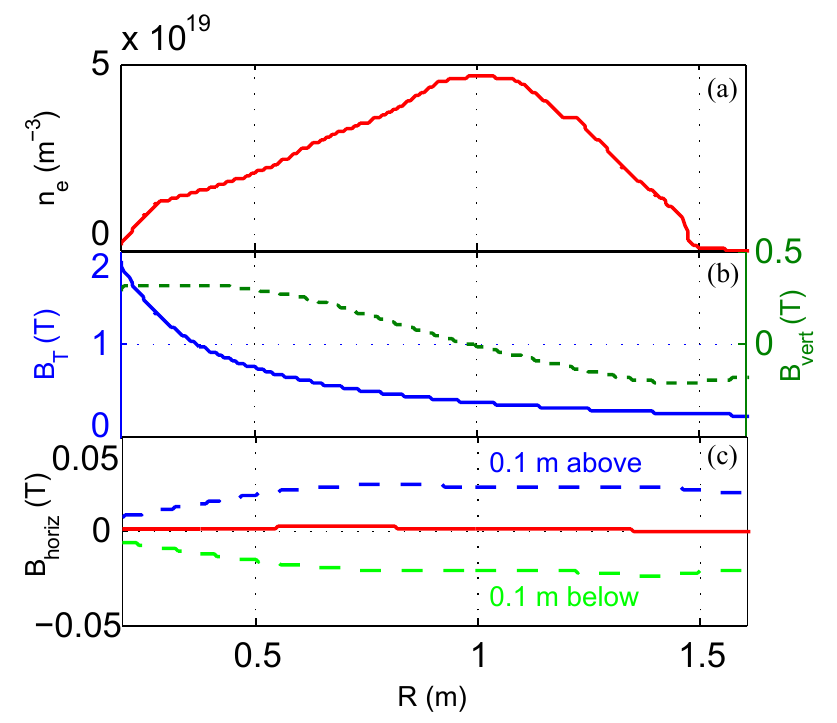}
 \caption{\label{fig:profiles}(a) Plasma density profile of base case for modeling. (shot \# $124764$, $0.325\ sec$). (b) Toroidal ({\color{blue}solid line}) and vertical ({\color{green}dashed line}) magnetic field along major radius in the mid-plane (they vary little with height near the mid-plane). (c) Horizontal (i.e.\ radial) magnetic field along major radius $0.1\ m$ above ({\color{blue} blue dashed line}), below ({\color{green} green dashed line}) and in the mid-plane ({\color{red} red solid line}).}
\end{figure}

Many of the modeling results discussed here are obtained for a $288\ GHz$ ($\lambda = 1.04\ mm$) probing microwave in a base case plasma (shot \# $124764$, $0.325\ sec$) that represents conditions likely to be encountered in NSTX. The base case is an L-mode plasma [Fig.~\ref{fig:profiles}] with a major radius of $R_0 = 0.85\ m$ and a minor radius $a = 0.67\ m$. The density profile is centrally peaked with a maximum of $n_0 = 4.7 \times 10^{19}\ m^{-3}$ at $R_{axis} = 1.0\ m$. The electron plasma and cyclotron frequencies are $61.4\ GHz$ and $10.5\ GHz$ on axis. The millimeter wavelength used is longer than typical for polarimetry systems, but it is a good compromise between two competing constrains for the base case. At longer wavelength the effects of the plasma on polarization are stronger, allowing for more sensitive measurement of magnetic fluctuations. At shorter wavelength, refraction becomes less significant.

\begin{figure}
\includegraphics[width = 7.0 cm]{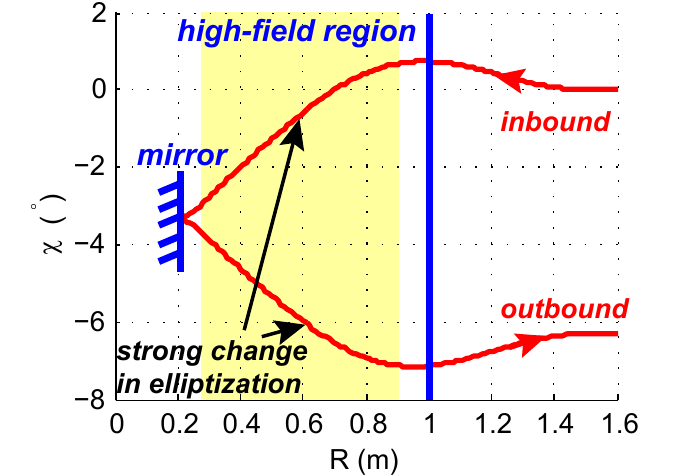}
 \caption{\label{fig:chi}Elliptization angle ($\chi$) evolution along chord in mid-plane for waves with horizontal linear polarization at launch (i.e.\ in toroidal direction). Vertical solid line indicates plasma center (i.e.\ peak density). The mirror is mounted on the center stack. ($f = 288\ GHz, \lambda = 1.04\ mm$)}
\end{figure}

The modeling shows that the magnitude of the elliptization angle increases most rapidly when the wave is in the high-field region ($R<R_{axis}$). [Fig.~\ref{fig:chi}] This stands in contrast with conventional tokamaks. Elliptization is sensitive to the strength of the perpendicular magnetic field (Eq.~\ref{eq:chi}), of which the toroidal magnetic field $B_T$ is a significant component. $B_T$ varies approximately inversely with major radius in both conventional tokamaks and spherical tori, but in spherical tori, the variation is much stronger because of their relatively low aspect ratio. For instance, in NSTX ($R_0/a \simeq 1.27$) $B_T$ varies from $0.2\ T$ at the outer edge ($R = 1.6\ m$) to $2\ T$ close to the center stack ($R = 0.2\ m$). [Fig.~\ref{fig:profiles}(c)]

\begin{figure}
\includegraphics[width = 7.0 cm]{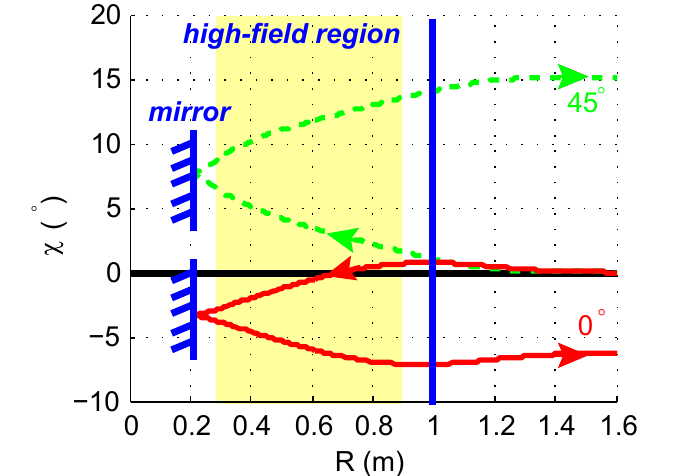}
 \caption{\label{fig:chi_0_45}Elliptization angle ($\chi$) evolution for horizontal ({\color{red} solid line}) and $45^{\circ}$ ({\color{green} dashed line}) linear polarization at launch along chord in mid-plane. Zero elliptization (i.e.\ linearly polarized) is highlighted on the grid. ($f = 288\ GHz, \lambda = 1.04\ mm$)}
\end{figure}

Modeling shows that the evolution of the elliptization depends strongly on the polarization direction in the high-field region. Fig.~\ref{fig:chi_0_45} shows the dramatically different elliptization evolution for two waves launched in the mid-plane with launch angles of $0^{\circ}$ and $45^{\circ}$. This dependence is expected since elliptization is strongly weighted to the high-field region and sensitive to $\psi$ (Eq.~\ref{eq:chi}). For a chord in the mid-plane, the polarization direction in the high-field region is determined by the launch angle since Faraday rotation is very weak there ($\vec{B_{\parallel}}$ is weak in the mid-plane).

\begin{figure}
\includegraphics[width = 7.0 cm]{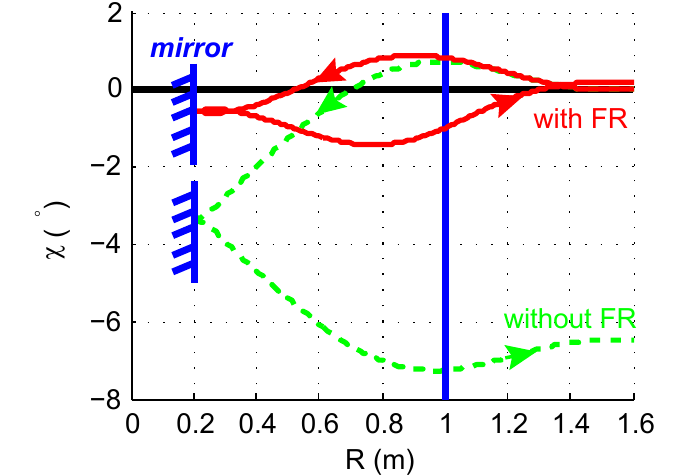}
 \caption{\label{fig:real_CM}Elliptization angle ($\chi$) evolution of wave with horizontal linear polarization at launch along a chord $0.1\ m$ above mid-plane with ({\color{red} solid line}) and without({\color{green} dashed line}) Faraday rotation. Faraday rotation is eliminated by setting $\vec{B_{\parallel}} = 0$ along chord). ($f = 288\ GHz, \lambda = 1.04\ mm$)}
\end{figure}

Of particular interest, the modeling shows that Faraday rotation can play a significant role in elliptization. Chords away from the mid-plane can have significant $\vec{B_{\parallel}}$, so Faraday rotation can substantially change the polarization direction of the wave before it enters the high-field region. Fig.~\ref{fig:real_CM} compares the elliptization evolution of a wave launched with horizontal linear polarization both with and without the influence of Faraday rotation. The final elliptization of the wave is very different for the two cases. The modeled chord is $0.1\ m$ above the mid-plane, where $|\vec{B_{\parallel}}|$ reaches a maximum of $0.024\ T$ along the chord. For the case without Faraday rotation,  $\vec{B_{\parallel}}$ is simply set uniformly to zero along the chord. The impact of Faraday rotation on elliptization identified here is a primary element of the interaction between the two effects.

\begin{figure}
\includegraphics[width = 7.0 cm]{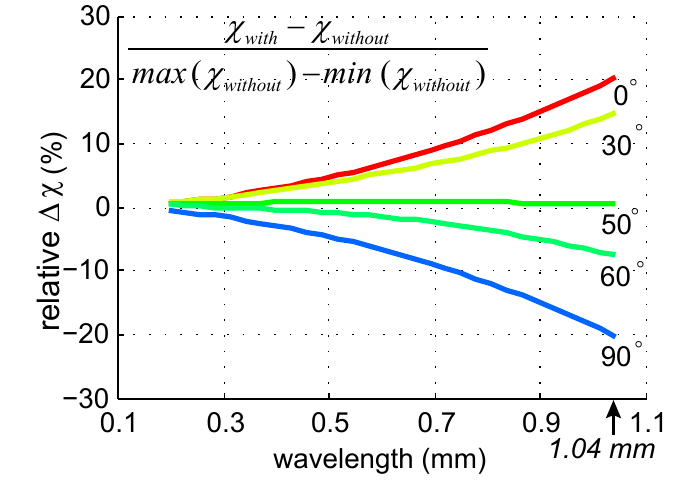}
 \caption{\label{fig:normal_chi}Difference between the final elliptizations with and without Faraday rotation ($\Delta \chi$) versus wavelength for different launch angles. $\Delta \chi$ is normalized by the difference between the maximum and minimum final values of $\chi$ without Faraday rotation.}
\end{figure}

Modeling shows a significant wavelength dependence in the strength of the impact of Faraday rotation on elliptization. Both the Faraday rotation and Cotton-Mouton effects are expected to become stronger with increasing wavelength (i.e.\ lower frequency). (Eq.~\ref{eq:chipsi_app}) However, it is not obvious whether the impact of Faraday rotation on elliptization should become more or less significant as the wavelength increases. To assess this, the change in the final value of $\chi$ caused by including Faraday rotation is calculated. [Fig.~\ref{fig:normal_chi}] The change $\Delta \chi$ is normalized by the difference between the maximum and minimum final values of $\chi$ without Faraday rotation. This normalization factor serves as a measure of the strength of the elliptization effect. Fig.~\ref{fig:normal_chi} shows that the relative effect of Faraday rotation on elliptization increases with wavelength.

The interaction complicates the interpretation of polarimetry measurements if both the Faraday rotation and Cotton-Mouton effects are large. This is an issue not only in the plasma modeled here, but whenever the wavelength is sufficiently long regarding the plasma conditions, such as in the planned $118\ \mu m$ polarimeter in ITER\cite{donne_poloidal_1999}. Under these conditions, the approximate expressions (integral forms of Eq.~\ref{eq:chipsi_app}) for the polarization rotation and elliptization are not appropriate. Also, the interpretation of an array of chord measurements used to characterize the equilibrium\cite{rommers_new_1996,brower_measurement_2002} becomes more complicated. The profile of final polarization direction versus chord impact parameter is affected by the interaction, leading to a change in both the zero crossing and slope.

\section{Summary and Conclusion}
This work models the evolution of EM wave polarization along major radial chords, with retroreflection on NSTX, using Mueller-Stokes calculus. Most of the modeling focuses on $288\ GHz$ microwaves launched with linear initial polarization in a base case plasma that represents conditions commonly encountered in NSTX. The modeling shows that the Cotton-Mouton effect is strongly weighted to the high-field region of NSTX. An interaction between the Faraday rotation and Cotton-Mouton effects is also clearly identified. Elliptization occurs when the wave polarization direction is neither parallel nor perpendicular to the local transverse magnetic field. Since Faraday rotation modifies the polarization direction during propagation, it must also affect the resultant elliptization. The Cotton-Mouton effect also intrinsically results in rotation of the polarization direction, but this is less significant to the modeling results presented here. The interaction identified here is shown to increase in significance with wavelength. Care has to be taken in interpreting the polarimetry measurement if both effects are large.

\section{Acknowledgment}
This work is supported by U.S.\ DOE Contract No.\ DE-FG02-99ER54527.

\bibliographystyle{aipnum4-1}
%\bibliographystyle{apsrev}
%\bibliography{2010_HTPD}

%Merlin.mbs v4.21 2009-07-09.
%

\end{document}